\documentclass{emulateapj}
\usepackage{times}

\newcommand{\msun}{\mbox{$M_{\odot}$}}

\shorttitle{Generating Hot Gas in Simulations of Disk-Galaxy Major Mergers}
\shortauthors{Cox et al.}

\begin{document}

\title{Generating Hot Gas in Simulations of Disk-Galaxy Major Mergers}

\author{T. J. Cox\altaffilmark{1}, Joel Primack\altaffilmark{1}, 
Patrik Jonsson\altaffilmark{2}, and Rachel Somerville\altaffilmark{3}}

\begin{abstract}
We report on the merger-induced generation of a shock-heated gas wind 
and formation of a remnant gas halo in simulations of colliding disk 
galaxies.  The simulations use cosmologically motivated initial conditions 
and include the effects of radiative cooling, star formation, stellar 
feedback and the non-adiabatic heating of gas.  The non-adiabatic heating, 
i.e. shocks, generated in the final merger forces gas out of the central 
region of the merger remnant and into the dark-matter halo.  We demonstrate 
that the amount of heating depends on the size of the progenitor disk
galaxy as well as the initial orbit the galaxies are placed on.  Based upon 
these dependencies, we motivate a possible recipe for including this effect 
in semi-analytic models of galaxy formation.
\end{abstract}

\altaffiltext{1}{Department of Physics, University of California, 
Santa Cruz, CA 95064, USA, tcox@physics.ucsc.edu}
\altaffiltext{2}{Department of Astronomy and Astrophysics, University of California, 
Santa Cruz, CA 95064, USA}
\altaffiltext{3}{Space Telescope Science Institute, 
3700 San Martin Drive, Baltimore MD, 21218, USA}

\section{Introduction}
\label{sec:intro}

Galactic winds are a common phenomenon associated with starbursting 
galaxies in the local \citep{M99,Heck00,St02} and high redshift 
\citep{Pet02,Adel03} universe.  It is generally assumed that feedback 
from massive stars, either stellar winds or supernova ejecta or both, 
power these outflows of hot gas.  The starburst origin of galactic winds 
in nearby galaxies is supported by the observed correlation between 
wind-induced mass loss and the star formation rate \citep*{M99,Ru02}. 
In addition to transporting mass and energy from the central star-forming 
region, galactic winds also transport metals \citep{Heck00}.  Depending 
on the energy in the winds and the depth of the halo potential well, 
outflows such as these may be responsible for heating and polluting the 
intergalactic medium.

Galactic winds likely play a significant role in galaxy formation and 
evolution.  As a source of energy, or {\it feedback}, galactic winds regulate 
the conversion of gas into stars.  In recent semi-analytic models of galaxy 
formation, winds (in addition to a UV background) are required to remove gas 
and suppress star formation in small mass halos, making the theoretical halo 
mass function consistent with the observed galaxy luminosity function 
\citep{S02,BenIII}.  Energy input due to galactic winds counteracts the 
efficient gas cooling in galactic halos - the so-called ``overcooling'' 
problem of theoretical models or simulations. Interestingly, not only do 
these models produce too many faint galaxies, but they typically produce too 
many bright galaxies as well.  Thus, overcooling is a problem in both low 
{\it and} high mass halos.  

While feedback from massive stars provides a physically reasonable and 
effective mechanism by which the low mass end of the halo mass function 
can be reconciled with the galaxy luminosity function, it is unclear that 
supernova-driven winds can provide enough energy to solve the overcooling 
problem for large halos.   \citet{Ben03}, using semi-analytic models, 
were able to match the bright end of the luminosity function only by 
assuming extremely efficient thermal conduction or by including galactic 
winds whose energy greatly exceeded that produced by supernovae.

It is clear additional sources of energy are required during the process of
galaxy formation and in this letter we suggest a possible candidate.  We 
demonstrate that simulations of disk-galaxy mergers generate shock-heated 
winds which can put gas with thermal energy up to 1 keV per baryon into 
the dark-matter halo.  These winds are {\it not} induced by supernovae, but 
instead caused by orbital energy which is transferred to the diffuse gas.  Since
major mergers are more common during the hierarchical formation of massive 
structures, energy input due to this process will preferentially act in more 
massive halos.  Further, the generation of hot gas due to the merging process 
suggests that we should be able to detect hot gas in nearby merging systems.

The structure of this letter is as follows: \S \ref{sec:sim} summarizes 
our simulation techniques and our initial conditions, \S \ref{sec:mergers} presents 
an overview of the galaxy merger process focusing on the conditions relevant
to generating hot galaxy winds, \S \ref{sec:outfm} gives properties of the outflow
material and presents a physically motivated fitting formula to capture our
results which we discuss in \S \ref{sec:disc}.

\section{Simulations}
\label{sec:sim}

We use GADGET \citep*{spGad} to simulate the merger between two 
identical copies of our model disk galaxy.  Based on the work of 
\citet{spEnt}, and adopting their terminology, we use the the 
conservative-entropy version of smoothed particle hydrodynamics (SPH),
although we remark that because we include shocks and radiative cooling 
entropy is not constant throughout the simulation.  Gas is assumed to 
be a primordial plasma and can cool, form stars and produce supernovae 
which release energy and return metals to the ISM.  Star formation is 
based upon the local dynamical time scale.  Stellar feedback is 
implemented similarly to \citet{sp00}; we define a feedback energy 
reservoir in which supernova feedback energy is stored.  This 
reservoir has a dissipative time scale longer than the cooling time and 
can provide pressure support in star-forming regions.  Temperature and 
pressure are calculated as the sum of the thermal and feedback 
reservoirs.  We note that we have varied our star formation and 
feedback assumptions and the results we report here are robust. Details 
of our methods are given in \citet{Cox}.

Star-formation and feedback parameters are adjusted so that an isolated
disk galaxy's star formation follows the empirical Schmidt law 
\citep{Kenn98}.  Once these parameters have been set, we collide two 
disk galaxies by fixing them on a prograde orbit defined by the 
eccentricity $e$ and the pericentric distance $R_{\rm peri}$, as 
determined from the initial energy and angular momentum.  One disk is 
in the orbital plane while the other is inclined by $30^\circ$.

We simulate mergers between four different identical pairs of progenitor 
disk galaxies.  The largest represents a disk of roughly 5x the mass of 
the Milky Way, which is our second largest progenitor galaxy.  We select 
parameters for the Milky Way from model A1 of \citet*{KZS} and for 
simplicity ignore the bulge component.  Our final two progenitors are 
dwarf galaxies of one-tenth and one-hundredth the mass of the Milky Way.  
Initial disk galaxy parameters are listed in Table~\ref{tab:ics}.

Each galaxy model consists of an exponential disk, of scale length 
R$_{\rm d}$ and gas fraction $f$, which resides in the center of a 
spherically symmetric NFW \citep*{NFW} dark matter halo.  The numerical 
construction of this system is described in \citet{sp00}.  We use 
180,000 total (40,000 gas) particles for our largest progenitor galaxy, 
70,000 (20,000) for our Milky Way-like galaxy and 30,000 (5,000) for 
each of the dwarfs.  GADGET smoothing lengths are 570 pc for dark matter 
particles and 140 pc for baryonic particles in our two larger systems 
while the dwarfs have smoothing lengths of 140 pc and 57 pc.  Identical 
copies of each galaxy model (as specified in Table~\ref{tab:ics}) are 
put on a variety of orbits which are listed in Table~\ref{tab:iorbits}. 
Each interaction was simulated until the merger was relaxed, typically 
700 Myr after the nuclei coalesce, and total simulation times were 
between 2-5 Gyr.

\begin{deluxetable}{lccccccc}
\tabletypesize{\scriptsize}
\tablecaption{Disk Galaxy Initial Conditions\label{tab:ics}}
\tablewidth{0pt}
\tablehead{
\colhead{Label} & \colhead{M$_{\rm vir}$} & \colhead{R$_{\rm vir}$} &
\colhead{c} & \colhead{$\lambda$} & \colhead{$m_d$} & \colhead{$f$} &
\colhead{R$_{\rm d}$}\\
\colhead{(1)} & \colhead{(2)} & \colhead{(3)} & \colhead{(4)} &
\colhead{(5)} & \colhead{(6)} & \colhead{(7)} & \colhead{(8)}
}
\startdata
5xMW & 700 & 499 & 11 & 0.05 & 0.04 & 0.5 & 10.6  \\
MW & 140 & 291 & 12 & 0.031 & 0.04 & 0.1 & 3.4  \\
Dwarf1 & 14 & 135 & 15 & 0.05 & 0.04 & 0.5 & 2.1  \\
Dwarf2 & 1.4 & 63 & 20 & 0.05 & 0.05 & 0.5 & 1.0  \\
\enddata 
\tablecomments{Column~1: Disk galaxy label. Column~2: Galaxy virial
mass in units of $10^{10}$ \msun;
Column~3: Virial Radius R$_{\rm vir}$, in kpc;
Column~4: Initial halo concentration c$\equiv$R$_{\rm vir}$/r$_{\rm s}$
where r$_{\rm s}$ is the NFW scale radius;
Column~5: Dimensionless spin parameter;
Column~6: Fraction of virial mass in an exponential baryonic disk;
Column~7: Fraction of exponential baryonic disk mass in a
collisional gaseous component;
Column~8: Disk scale length R$_{\rm d}$, in kpc.
}
\end{deluxetable}

\begin{deluxetable*}{lcc}
\tabletypesize{\scriptsize}
\tablecaption{Galaxy Collision Orbits\label{tab:iorbits}}
\tablewidth{0pt}
\tablehead{
\colhead{Label} & \colhead{R$_{\rm start}$} &
\colhead{(R$_{\rm peri}$, $e$)}
}
\startdata
5xMW   & 429 & (1.7, 0.99), (7.1, 1), (21.4, 1), (35.7, 0.9), (35.7, 1), (85.7, 0.6) \\
MW     & 286 & (0.1, 0.99), (7.1, 1), (21.4, 0.9), (35.7, 0.8), (178.6, 0) \\
Dwarf1 & 143 & (1.1, 1), (2.9, 1), (5.7, 1), (10, 1), (15.7, 1), (21.4, 1), (31.4, 1) \\
Dwarf2 & 57  & (0.6, 1), (1.1, 1), (2.4, 1), (4.1, 1), (6.4, 1), (8.9, 1), (12.9, 1) \\
\enddata
\tablecomments{Column~1: Disk galaxy label. Column~2: Initial galaxy
separation R$_{\rm start}$, in kpc;
Column~3: The pericentric distance R$_{\rm peri}$, in kpc, and
the eccentricity of each orbit simulated.}
\end{deluxetable*}

\section{Galaxy Mergers}
\label{sec:mergers}

\begin{figure}
\plotone{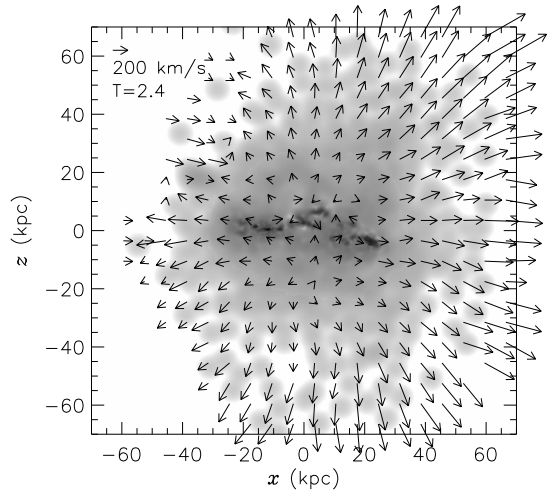}
\caption{
Velocity field of the gas overlaid on a gray scale map of the 
gas density in a 7 kpc slice through a plane perpendicular to 
the orbital plane at 2.4 Gyr after the start of the simulation
and 200 Myr before the final merger.  The length of the arrow 
in the upper left corner represents 200 km/s.  From this view
the coherent wind motion is clearly visible as it escapes the
high-density merging region out of the orbital plane, i.e.
moving along the path of least resistance.
}
\label{fig:vel-field}
\end{figure}

The morphology and induced star formation of gaseous galaxy collisions 
has been discussed  in a number of papers \citep{BH:1996,MH:1996,sp00}.  
Here we briefly emphasize the aspects of the merger process relevant for 
our purpose.  The simulation begins with a wide separation between the 
two galaxies and they are essentially unaffected by each other's presence.  
As the galaxies reach pericenter the disks become tidally distorted and a 
small burst of star formation ensues.  At this point dynamical friction 
has extracted orbital energy from the originally unbound orbit but the 
galaxies are still able to briefly separate before they finally merge. At 
the final merger the two disks become effectively destroyed and we are 
left with a spheroidal-looking object.

We find that the total fraction of gas converted to stars during the 
simulations ranged from 68\% to 92\% in line with previous studies 
\citep{MH:1996,sp00}.  While the star formation is generally efficient at 
converting gas to stars, we note that the bursts of star formation during 
the final merger are less efficient than in previous work.  In \citet{Cox} 
we show that this results from using the conservative-entropy version of 
SPH.  Each orbit has a unique star formation history which dictates how 
much star formation occurs quiescently or during tidal triggered episodes, 
but we note that the total gas consumption is only weakly correlated with 
orbit.

At the time of the final merger between the disk galaxies, the majority of 
remaining gas is cold and of relatively low density.  Most of this gas was 
in the outer portions of the original disks.  Some of this gas was ejected 
in long tidal tails during the first encounter of the two disk galaxies, 
but by the final merger much has fallen back to its parent galaxy.  As the 
two primarily cold gas disks attempt to follow their dominant dark halos 
into the final merger, the collisional nature of the gas begins to dominate 
and a large amount of gas shocking occurs.  These shocks can heat the gas 
to temperatures between $10^6$ and $10^7$K.

The hot gas then expands adiabatically and finds little resistance out of 
the orbital plane as shown in Figure~\ref{fig:vel-field}.  Gas collectively 
moves at speeds $\ge$ 200 km/sec, and quickly fills the dark-matter halo.  
It is often the case that there are several episodes of gas ejection 
similar to Figure~\ref{fig:vel-field} due to the multiple shock events as 
the nuclei oscillate and finally merge.  Since we have not included a 
gaseous halo in our initial conditions we do not capture the interaction of 
the wind with the ambient gas within the original halo.

Within the gaseous component, the final merger corresponds to a transfer 
of bulk kinetic energy to thermal energy, indicative of shocks.  While the 
amount of energy generated by supernovae associated with star formation 
during the course of the merger is quite large, its contribution is 
localized to the high density regions and spread out over time.  The 
supernovae feedback energy contributes a negligible amount to the energy 
and entropy gain in the lower density gas at the final merger.  During the 
final merger the majority of the entropy gain is due to low-density gas, 
which is shock heated while never (or only briefly) reaching star-forming 
densities.  This emphasizes that the winds reported here are primarily due 
to shocks and {\it not} star formation.  One might expect the thermal 
energy to be efficiently radiated away, but in practice the gas quickly 
accelerates out of the central region to densities where the cooling time 
ranges from $\geq$ 300 Myr to several Gyr.

We verified that resolution effects were not limiting our ability to
capture the shock heating during the galaxy mergers by resimulating the 
MW(R$_{\rm peri}$=7.4,$e$=1) orbit with a factor of 10 higher and lower
mass resolution.  The amount of hot gas in the merger remnant and the 
average energy of this gas varied by less than 3\% in these resimulations, 
emphasizing that we are not limited by resolution.  We speculate that 
cosmological simulations do not see shock heating in mergers because they 
efficiently form stars from all cold gas inside dark matter halos long 
before the dark matter halos finally merge.  This will be investigated in 
a future paper.

The ``phase difference'' \citep{NW:1993} between the dominant collisionless 
center of mass and the gas during the final merger is dictated by the orbits 
the initial galaxies are placed on.  Nearly radial orbits tend to have their 
nuclei oscillate through each other during the final merger providing for 
ample phase lags and hence ample shock heating.  Conversely, circular orbits 
demonstrate very little gas heating during the final merger because the 
nuclei spiral around each other as the galaxy centers coalesce.  In this 
case gas from each disk quiescently `slides' around the other disk's gas and 
little shocking takes place.

\section{Hot Gas in Merger Remnants}
\label{sec:outfm}

\begin{figure}
\plotone{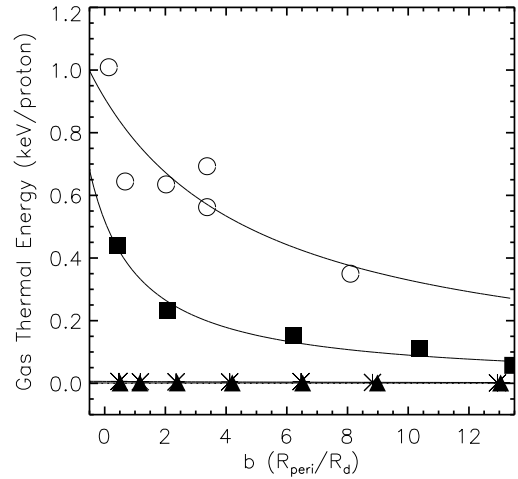}
\caption{
Remnant gas energy in keV per proton as measured 500 Myr after
the merger between two disk galaxies plotted as a function of
the orbital pericentric distance R$_{\rm peri}$ divided by the
original disk scale length R$_d$.  Open circles are our the
progenitor disk galaxy 5xMW, the largest progenitor disk galaxy,
solid squares are MW disks, the asterisk are Dwarf1 and solid
triangles are Dwarf2.  Solid lines are fits to the data using
equation~(\ref{fittingf}) and described in the text.
}
\label{fig:evmp}
\end{figure}

In Figure~\ref{fig:evmp} we detail the thermal energy per proton of all 
gas in each merger remnant.  There is a trend for orbits with smaller 
impact parameters, here called $b$ and defined as 
R$_{\rm peri}/{\rm R}_{\rm d}$, to generate gas with more thermal energy.  
Since our initial conditions did not provide any low density medium 
representing ambient gas we are not capturing the inevitable process by 
which the out-flowing gas can thermalize any of its kinetic energy.  We 
do not show this kinetic energy in Figure~\ref{fig:evmp} since there are 
additional uncertainties with regard to energy loss due to radiation or 
the production of cosmic rays, but we remark that the wind's kinetic 
energy is approximately 10-50\% of its thermal energy.

From energy considerations alone we can attain an approximation of the 
gas thermal energy plotted in Figure~\ref{fig:evmp}.  While the initial
kinetic energy and total orbital energy are dependent on the assumed
eccentricity and pericentric distance, the initial potential energy per
unit mass is
\begin{equation}
\frac{GM_{\rm vir}}{{\rm R}_{\rm start}} = 
0.22~{\rm keV}/{\rm proton}
\biggl(\frac{M_{\rm vir}}{10^{12} \msun}\biggr)
\biggl(\frac{200 kpc}{{\rm R}_{\rm start}}\biggr),
\label{pe0}
\end{equation}
independent of the orbit.  During the merger this energy is absorbed by 
the galaxies, consistent with Figure~\ref{fig:evmp}.

For a more complete understanding of the origin of Figure~\ref{fig:evmp}
we must consider how shocks thermalize kinetic energy.  The initial disk 
gas is cold (temperature $\sim 10^4$~K) prior to shock heating and the 
sound speed is $\leq$ 15 km/sec.  The orbits we assume dictate the 
typical velocity with which the galaxies interact to be $\geq$ 
100 km/sec and hence the gas in each progenitor disk will undergo strong 
shocks (Mach number $\gtrsim$ 10) as the disks collide during the final 
merger.  The Rankine-Hugoniot jump conditions in the strong shock regime
give
\begin{equation}
kT_{\rm final} \approx \frac{3}{16}mv_{\rm initial}^2,
\label{shock}
\end{equation}
where T$_{\rm final}$ is the temperature of the gas after the shock, and 
$mv_{\rm initial}^2$ is the initial kinetic energy of gas. In physical 
terms the jump condition states that the initial kinetic energy is 
transfered to gas thermal energy via the strong shock with almost 50\% 
efficiency.  Moreover, the strong shock only occurs when the progenitor 
disk galaxies pass through each other rather than orbit around each other, 
and thus we reason it is the initial kinetic energy in the radial direction 
which gets converted to heat because of shocks, i.e. $v_{\rm initial} 
\sim v_{\rm radial}$.

It is straightforward to determine the initial kinetic energy along the 
line connecting the centers of the initial disk galaxies as
\begin{equation}
v_{\rm radial}^2 \sim GM_{\rm vir} \frac{e^2}{(1+e)}\frac{1}{R_{\rm peri}},
\label{ker}
\end{equation}
where R$_{\rm peri}$ and $e$ are the pericentric distance and orbit 
eccentricity, respectively.  We see that the initial radial kinetic energy 
is inversely proportional to the pericentric distance, and hence initial 
orbits with large impact parameters should have much less gas shocking and 
thus less hot gas in their merger remnant, as seen in Figure~\ref{fig:evmp}.

Based upon the above assumptions we propose a model by which we can 
capture the general features of generating hot gas by disk galaxy major 
mergers.  Primarily the generation of hot gas is inversely proportional 
to the impact parameter $b$ defined above.  We find that the following 
simple formula accurately captures this relationship,
\begin{equation}
kT_{\rm final}/{\rm proton} = 
\frac{\epsilon_{\rm merg}}{b + {\rm R}_{\rm d}/{\rm R}_A},
\label{fittingf}
\end{equation}
where $\epsilon_{\rm merg}$ is a constant with units energy per proton, $b$ 
is the impact parameter defined above, R$_{\rm d}$ is the disk radial scale 
length and R$_A$ is a distance which acts as a fudge factor to scale the 
heating dependence on the disk size.  This formula is fit to the data using 
a least squares fitting routine and displayed in Figure~\ref{fig:evmp}.  For 
mergers of the 5xMW and MW progenitor disk galaxies we find 
$\epsilon_{\rm merg}$ scales with the mass as expected from 
equation~(\ref{ker}), its value can be approximated as 
\begin{equation}
\epsilon_{\rm merg} \sim 1 {\rm keV}/{\rm proton}
	  \biggl(\frac{M_{\rm vir}}{10^{12} \msun}\biggr),
\label{fittedA}
\end{equation}
and R$_A$ is roughly 1.18 kpc.  Fits for the smallest progenitors, Dwarf1 
and Dwarf2, are not well approximated by this formula as the thermal 
energy of all remnants was approximately 3 eV/proton and 1 eV/proton, 
respectively.  We expect a fair amount of scatter due to varying 
gas distributions and gas mass fractions in the progenitor disk galaxies. 
More simulations must be performed over a wide range of initial 
conditions to determine if equation~(\ref{fittingf}) or~(\ref{pe0}) better 
captures the full thermalization of orbital energy due to the merger.

Very little hot gas is energetic enough to escape the potential 
well of the halo, and the majority comes into equilibrium with a 
profile reasonably well represented by a spherical $\beta$-profile 
\citet{Eke98}.  The entropy per proton generated via the merger 
process is $\leq$ 400 keV cm$^2$ and follows a trend very similar to
that seen in Figure~\ref{fig:evmp}.

\section{Discussion}
\label{sec:disc}

Hydrodynamic simulations of colliding disk galaxies demonstrate the
ability to generate a significant amount of hot gas from shocks which 
abundantly occur in the central several kpc region during galaxy 
collisions.  As the centers of disk galaxies coalesce, collisional gas 
cannot inter-penetrate and gets heated while attempting to follow the 
potential well dominated by the collisionless dark matter and stellar 
components.  We note that any isothermal gas assumption used to simplify 
the treatment of gas \citep{BH:1996,MH:1996} would not reproduce the 
process we discuss here.

In general, hot gas generated by galaxy major mergers could be an 
important source of energy during the process of galaxy formation.  Not 
only do radial orbits pump energy and entropy into the gas but they 
provide an effective mechanism to recycle moderately dense cold gas 
into diffuse hot gas, providing a feedback loop for energy, metals and 
mass that correlates with the hierarchical build-up of any galaxy.
In reality the process we describe here may be augmented by 
contributions from stellar winds, supernovae, and AGN which will 
increase the total gas heating.  The remnants we report here have 
cooling times of several Gyr and could provide an enriched medium with 
which a gas disk could subsequently be formed.

Our results suggest that any galaxy system which has been involved in 
a major merger should have hot gas residing throughout its dark matter 
halo.  Additionally, the temperature of this gas, at a fixed time after 
the final merger, should correlate with the mass of the progenitor disk 
galaxies and their merger orbit.  Further work will investigate if the 
profiles in our merger remnants resemble the extended gas distributions 
found in relaxed elliptical dominated groups as found by \citet{mz98} or 
if the outflowing gas is similar to the diffuse gas found in ongoing mergers 
such as Arp220 \citep{McD03} and `The Antennae' \citep{Fab01}.

\acknowledgments We thank Volker Springel for help with GADGET and useful
comments on this research.  This research used resources of the National
Energy Research Scientific Computing Center (NERSC), which is supported by 
the Office of Science of the US Department of Energy.

\end{document}